\RequirePackage{fix-cm}
\documentclass[smallextended]{svjour3}       
\smartqed  
\usepackage{graphicx}
\usepackage{amsmath}
\usepackage{amsfonts}
\usepackage{braket}
\usepackage{bm}
\usepackage{caption}
 \usepackage{mathptmx}      
\DeclareMathOperator{\Tr}{Tr}
\DeclareMathOperator{\CR}{CR}
\DeclareMathOperator{\IPT}{IPT}
\DeclareMathOperator{\CNOT}{CNOT}
\DeclareMathOperator{\NOT}{NOT}

 \journalname{Quantum Information Processing}
\begin{document}

\title{A linear time quantum algorithm for 3SAT using irreversible quantum operations
}


\author{Zachary B. Walters
}


\institute{Z. Walters \at
                Lawrence Livermore National Lab \\
                7000 East Avenue \\
                Livermore, CA 94551 \\
              \email{walters24@llnl.gov}           
}

\date{Received: date / Accepted: date}

\maketitle

\begin{abstract}

        The Deutsch model of quantum computation is extended to allow for thermodynamically irreversible operations by allowing the system of interest to interact with an outside reservoir.  A set of irreversible logical error correction superoperators are constructed which allow the rapid concentration of probability from an exponentially large search space into a small number of logically defined states.  These capabilities are used to construct a linear time solution algorithm for the NP complete problem 3SAT.  
\end{abstract}

\section{Introduction}
\label{sect:introduction}

The idea that the laws of quantum mechanics might be applied to the solution
of computational problems is usually attributed to Richard
Feynman\cite{feynman1982simulating}, who observed that the computational
resources required to simulate many quantum systems grow exponentially with
respect to system size.  If a system may be regarded as ``calculating'' its
own evolution, it follows that some principle of quantum mechanics allows the
solution of some types of problems more quickly than is possible using
classical computation.  This insight has been borne out by subsequent
research, showing the existence of some problems that can be solved
exponentially faster by quantum than by classical
computation\cite{simon1997power}.  However, despite its power, Feynman's
insight is not prescriptive.  It gives no answers as to the class of
computational problems that might be accelerated by the use of quantum
mechanics, nor to the set of quantum mechanical behaviors that might be
usefully applied to computation.

To date, the most successful answer to these questions has been provided by
Deutsch\cite{deutsch1985quantum}.  In Deutsch's model, a quantum system
initially in pure state $\ket{\psi_{i}}$ is acted upon by a unitary operation
$U$ to produce a final state $\ket{\psi_{f}}= U \ket{\psi_{i}}$, which may
then be measured to probabilistically yield an answer to some computational
problem of interest.  Deutsch showed that a small set of one- and two bit
unitary operators, called a universal set, can be combined to yield an
arbitrary unitary operation on N bits.  A quantum algorithm then corresponds
to a unitary transformation, which is the product of the unitary
transformations making up the individual steps of the algorithm.  Since
Deutsch's original work, a number of competing methods have been developed for
implementing arbitrary unitary transformations, including the quantum gate
array\cite{deutsch1989quantum} \cite{feynman1986quantum}, adiabatic quantum
computation\cite{farhi2000quantum} \cite{aharonov2008adiabatic}, topological quantum
computation\cite{freedman2003topological}, and one-way quantum
computation\cite{raussendorf2001one}.  As each of these methods follows the
general outline of unitary transformations mapping wavefunctions to
wavefunctions which was originally set by Deutsch, they will be collectively
referred to as the Deutsch model of quantum computation.

Although the Deutsch model is universal in the sense that it can produce an
arbitrary unitary transformation mapping one wavefunction to another, it is not
universal in the sense of allowing any transformation of a quantum system that
is allowed by the laws of physics.  In general, a quantum system is not
described by a wavefunction but rather by a density matrix, and the evolution
of a quantum system is trace preserving but not unitary.  The restriction to
unitary transformations was a conscious design choice made by
Deutsch\cite{deutsch1985quantum}, who sought to generalize the theory of
reversible classical computation.  The success of this choice speaks for
itself; however, this does not preclude making different choices in pursuit of
different goals.

Physically, the restriction to unitary transformations of pure wavefunctions
is very limiting.  A unitary transformation of a quantum system is always
thermodynamically reversible -- it cannot change the entropy of the system it
acts upon.  This limitation is not echoed by the laws of physics, which allow
the entropy of an open quantum system to 
change when it interacts with the outside world.  Here it is
interesting to note that the restriction to reversible dynamics is not a
fundamental limitation in the realm of classical computation, where a
reversible computer can efficiently simulate the behavior of its irreversible
counterpart\cite{bennett1973logical}.  In contrast, a number of well-known
results involving the Deutsch model of quantum computation suggest that the
requirement of unitarity does restrict the behavior of a quantum computer.
The no-cloning theorem\cite{wootters1982single}, Deutsch's argument that a
quantum computer does not reduce the expected running time for a random
classical algorithm\cite{deutsch1985quantum}, and Bennett et al's proof that,
relative to a random oracle, a quantum Turing machine cannot solve class NP in
polynomial time\cite{bennett1997strengths}, as well as many
others\cite{nielsen2002quantum}, all rely heavily on the properties of unitary
transformations.   

Previous treatments of irreversible quantum algorithms include
\cite{verstraete2009quantum} \cite{kraus2008preparation}, using dissipation to drive
the system toward a desired steady state, 

This paper investigates the theory of irreversible quantum computation.  Here a
system of interest is allowed to interact with an outside reservoir, thereby
changing a closed quantum system to one which is open.  Unitary evolution of
system plus reservoir, combined with tracing over the states of the reservoir,
yields irreversible evolution for the system of interest.  Irreversible
operations allow for the concentration of probability from multiple initial
states into the same final state, thereby decreasing the entropy of the system
in a way which has no reversible counterpart.  A set of logical error
correction superoperators is defined, which irreversibly transfer population
from states that fail various logical error criteria to states that satisfy
them.  Applying these superoperators in series then allows the concentration of
probability from exponentially many initial states into a small number of
logically determined final states.  This is used to construct a linear time
solution algorithm for the NP complete problem 3SAT.

Section \ref{sect:open_quantum_systems} of this paper gives a brief
introduction to open quantum systems and the distinction between reversible and
irreversible quantum processes.  Section \ref{sect:logical_error_correction}
extends the Deutsch model to include irreversible operations, and develops a
system of logical error correction which uses incoherent population transfer to
decrease the population of states which fail to exhibit various logical
properties which are required for a valid solution.  Section
\ref{sect:running_time} analyzes the running time for the algorithm developed
in Section \ref{sect:logical_error_correction} when applied to 3SAT.

\section{Open quantum systems}
\label{sect:open_quantum_systems}

The theory of open quantum systems is well developed
elsewhere\cite{breuer2002theory} \cite{schlosshauer2007decoherence}.  For the
current purposes, the topic of interest is the connection between unitarity,
entropy, and thermodynamic work.

Although the basic laws of physics are time reversible, irreversible dynamics
arise when an isolated system is allowed to interact with the outside world,
here referred to as a reservoir.  The state of the reservoir may be
experimentally inaccessible, as is the case when a quantum system interacts
strongly with its surroundings, or an experimenter may simply partition an
isolated quantum system, treating one component as the system of interest and
the remainder as the reservoir.  Writing the density matrix for the system plus
reservoir as
\begin{equation}
    \rho_{SR} = \sum_{S,R,S',R'} \ket{S,R}\rho_{S,R; S', R'}\bra{S',R'},
\end{equation}
where $\ket{S,R}$ gives the combined state of system and reservoir and
$\rho_{S,R; S',R'}$ is a single element of the density matrix, the reduced
density matrix for the system alone is obtained by tracing over the states of
the reservoir
\begin{equation}
	\rho_{S} = \Tr_{R} \rho_{SR}.
\end{equation}
Note that although the trace operation may sum over an exponentially large
number of reservoir states, it does not require an exponentially large number
of operations to do so.  To the contrary, any measurement of system properties
implicitly involves a sum over all states of the reservoir which are consistent
with the measured value.  To avoid a trace, a measurement must resolve the
state of the reservoir as well as the state of the system of interest.



\subsection{Entropy and Work}

Tracing over the states of the reservoir results in a physically significant
departure from the Deutsch model.  In the Deutsch model, the system is isolated
from its surroundings, and time reversibility requires that the evolution of a
closed system be unitary.  In contrast, for an open system it is the combined
evolution of system plus reservoir which must be unitary.  After tracing over
the states of the reservoir
\begin{equation}
	\rho'_{S} = \Tr_{R} U_{SR} \rho_{SR} U^{\dag}_{SR}.
\end{equation}
the evolution $\rho_{S} \rightarrow \rho'_{S}$ of the system's reduced density
matrix will in general be nonunitary.

As the size of the reservoir increases, the number of states that are summed
over in the trace operation grows exponentially.  For most purposes, it is
more convenient to treat the entropy, or the logarithm of the number of states
being summed over, than the number of states itself. The Von Neumann entropy
of a quantum system is given by \cite{breuer2002theory}
\begin{equation}
    S = -\Tr[\rho \log(\rho)]/\log(2),
    \label{eq:vonneumanns}
\end{equation}
and is unchanged by a unitary operation mapping $\rho \rightarrow U\rho
U^{\dag}$.  Here $\log(2)$ is a normalization factor giving entropy the units
of bits, so that a completely disordered density matrix with equal population
in $2^{N}$ states has an entropy of $N$ bits.  Thus, the Deutsch model of
quantum computation is thermodynamically reversible -- it cannot change the
entropy of the system being acted upon.  An operation that changes the entropy
of a system is thermodynamically irreversible, and performs work on the system
equal to 
\begin{equation} 
    W = -T \Delta S, 
    \label{eq:work}
\end{equation} 
where $T$ is the temperature.  In this context, the temperature is unimportant
and may be set to 1.  However, the ability to perform work on the system is a
fundamental distinction between the Deutsch model and the current approach.


\subsection{Relation to computation} 

Although reversibility and entropy are usually understood from a physical
perspective, they have computational implications as well.  From a physical
perspective, a quantum computer is a machine for concentrating the probability
distribution of a quantum system into one or more states which satisfy logical
criteria.  The quantum system may begin in a pure wavefunction with entropy
zero, a completely disordered density matrix with entropy $N$ bits, or any
point in between.

The diagonal elements of the density matrix at any point during the execution
of the algorithm represent a probability distribution over a search space
consisting of states which have not yet been eliminated from consideration as
potential solutions.  The entropy is then the logarithm of the total phase
space volume occupied by the system at a given point in time, where a volume of
$\hbar^{3}$ equates to a single state.

In this picture, selectively measuring the state of the system is equivalent to
randomly sampling a point within the phase space volume that it occupies, in
hopes that the point lies within the subvolume which corresponds to valid
solutions.  The probability of success depends on three quantities: the total
volume being sampled from, the volume corresponding to correct solutions, and
the relative enhancement of probability in the volume of correct solutions
relative to the volume at large.

Consider a computational problem with $N_{V}$ valid solutions $s^{*}$.  The
total probability of obtaining a correct solution by measuring the state of the
system bits is given by
\begin{equation}
        P_{\text{correct}} = \sum_{s*} \rho_{s^{*} s^{*}}.
\end{equation}
Define 
\begin{equation}
        f_{v}= \frac{P_{\text{correct}}}{N_{v}/2^{S}},
\end{equation}
where $S$ is the system entropy given by Eq. \ref{eq:vonneumanns} with
$T=1$,  to be the relative enrichment of the solution states -- the ratio of
the probability concentrated within the solution states to the fraction of the
phase space volume which they occupy.  Then the probability of obtaining a
correct solution can be rewritten as
\begin{equation}
        P_{correct}= \frac{N_{v}f_{v}}{2^{S}}.
        \label{eq:volfrac}
\end{equation}

By definition, a reversible algorithm leaves $S$ unchanged, and must operate by
changing $f_{v}$, the relative enrichment of the valid solutions.  In contrast,
an irreversible algorithm may change $S$ as well as $f_{v}$ by doing work on
the system.  Reducing the system's entropy while leaving $f_{v}$ unchanged
corresponds to an exponential decrease in the phase space volume being sampled,
with a corresponding increase in the likelihood of success.  As will be seen,
this exponential speedup allows for the rapid solution of problems long
considered computationally intractable.

\section{Irreversible solution of logical problems}
\label{sect:logical_error_correction}

As seen in Eq. \ref{eq:volfrac}, a major advantage of irreversible algorithms
over their reversible counterparts is the ability to exponentially increase the
probability of measuring a valid solution by doing work on the system.  In
order to achieve this speedup, it is necessary to decrease the entropy by
concentrating probability into fewer states without decreasing the relative
enrichment of the solution states.

The approach which is taken in this paper is to develop a set of logical error
correction superoperators.  These operators decrease the population of states
which fail to satisfy various error criteria by incoherently transferring
population to states which satisfy them.

Let an error criterion $C$ be some Boolean formula indicating that a state is
not an acceptable answer, and let a logical problem $L = \{C\}$ be a set of
error criteria that must be satisfied.  Then a superoperator $E^{(C)}$ can be
called error correcting if it monotonically decreases the population of of
states failing clause $C$ for any choice of input density matrix $\rho$.  An
error correction superoperator may be said to be strongly error correcting if
all transferred population flows from states which fail the clause to states
which satisfy it, and weakly error correcting if some population is transferred
from some state $\ket{\alpha}$ which fails the clause to some other state
$\ket{\alpha'}$ which also fails the clause.  By definition, an error
correcting superoperator may not transfer any population from a state which
satisfies the clause to one which does not satisfy it.

This section constructs a set of error correction superoperators whose error
criteria correspond to the logical clauses in a 3SAT problem.  It first extends
the Deutsch model to allow for irreversible operations, then uses the expanded
model to define an irreversible population transfer superoperator.  This
superoperator is then modified to create an error correcting superoperator by
making the transfer of population conditional on the satisfaction of a logical
clause.  Successively applying these error correction superoperators will then
concentrate probability in those states which never fail a clause -- the
solutions to $L$.

\subsection{Extension of the Deutsch model}

The Deutsch model may be extended to treat open systems by introducing two
new operations.  Define an insertion operation to be the introduction of a new
bit, whose state at the time of insertion is separable from that of the system
of interest
\begin{equation}
	\mathfrak{I}_{\rho_{R}}\rho_{S} = \rho_{S+R}=\rho_{s} \otimes \rho_{R},
\end{equation}
where $\otimes$ indicates an outer product.  In this way, an insertion
operation maps a density matrix for $N$ bits onto a density matrix for $N+1$
bits.

A deletion operation removes a bit from the system of interest, mapping a
density matrix for $N$ bits onto a density matrix for $N-1$ bits.  This can be
accomplished in multiple ways.  Deletion by partitioning simply sets the bit
aside, to be ignored for the rest of the calculation.  All measurements of the
state of the system will then trace over the state of the partitioned bit.
Deletion by measurement projects the state of some bit onto the eigenstates of
some measurement operator.  Following Von Neumann, deletion by measurement can
be further subdivided into two types depending on whether the measured value is
retained.  If so, the measurement is selective, and measuring eigenvalue
$\alpha$ changes the system density matrix according to 
\begin{equation}
	\rho' = \mathfrak{M}_{\alpha} [\rho] =\sum_{i} \ket{\alpha_{i}}\bra{\alpha_{i}} \rho
	\ket{\alpha_{i}}\bra{\alpha_{i}},
\end{equation}
where the summation goes over all eigenstates having eigenvalue $\alpha$.
If the measured value is not retained, the measurement is nonselective, and
the summation goes over all eigenstates of the measurement operator
\begin{equation}
	\rho' = \mathfrak{N}_{\alpha}[\rho]=\sum_{\alpha,i} \ket{\alpha_{i}}\bra{\alpha_{i}} \rho
	\ket{\alpha_{i}}\bra{\alpha_{i}}.
\end{equation}

\subsection{Irreversible population transfer superoperator}

The use of insertion and deletion operations allows for irreversible transfer
of population from undesired states to desired states.  A simple example is
given by a two state system, in which population is transferred from state
$\ket{1}$ to state $\ket{0}$.  We first define a controlled rotation operator
\begin{equation}
  \CR^{c \rightarrow t}(\theta)=R^{t}(\theta/2) \exp[i \sigma^{c}_{z}
	\sigma^{t}_{y} \theta/4] = \left( \begin{array}{cccc} \cos(\theta/2) &
		-\sin(\theta/2) & 0 & 0 \\ \sin(\theta/2) & \cos(\theta/2) & 0 & 0 \\ 0 & 0 & 1
	& 0 \\ 0 & 0 & 0 & 1 \end{array} \right),
\label{eq:CR}
\end{equation}
where 
\begin{equation}
    R^{t}(\theta) = \exp[i \sigma_{y} \theta]= \left( \begin{array}{cc}
\cos(\theta) & -\sin(\theta) \\ \sin(\theta) & \cos(\theta) \end{array}\right)
\end{equation}
is an operator which rotates bit $t$ about the y axis, where $\ket{\pm y}=
(\ket{0} \pm i \ket{1})/\sqrt{2}$ and $\sigma_{x}$, $\sigma_{y}$, and
$\sigma_{z}$ are Pauli matrices.  The controlled rotation operator causes the
state of target variable $t$ to be rotated by angle $\theta$ when the
projection of the control bit onto control axis $\hat{z}$ is positive and left
unchanged if the projection is negative.  To reduce confusion, this paper will
follow the convention that when an arrow is used in the superscript for some
operator, the bit or bits to the left of the arrow control the evolution of the
bit or bits to the right.

Incoherent population transfer can be accomplished by first entangling the
state of the system with the state of an inserted reservoir bit, then using
the state of the reservoir bit to control the evolution of the system before
ultimately tracing over the state of the reservoir bit:  
\begin{equation}
	\IPT^{s}(\theta) = \Tr_{r}\mathfrak{N}^{r}_{x}\CR^{r \rightarrow s}(2 \theta) \CNOT^{s \rightarrow
	r} \mathfrak{I}_{ \ket{r}=\ket{0}},
	\label{eq:IPT}
\end{equation}
where
\begin{equation}
    \CNOT = \left( \begin{array}{cccc}
        0 & 1 & 0 & 0 \\
        1 & 0 & 0 & 0 \\
        0 & 0 & 1 & 0 \\
        0 & 0 & 0 & 1
    \end{array} \right)
\end{equation}
flips the state of the target bit if the state of the control bit is 1, and
leaves it alone otherwise.  Here the order of operations is right to left,
following the usual physics convention, and the reservoir bit is initially
inserted in state
$\ket{0}$.

The behavior of the population transfer superoperator defined in Eq.
\ref{eq:IPT} can be understood by following an arbitrary two state density
matrix through the individual steps.  Beginning with
\begin{equation}
	\rho= \left(
	\begin{array}{cc}
		\rho_{0 0} & \rho_{0 1} \\
		\rho_{1 0} & \rho_{1 1}
	\end{array}
	\right) 
\end{equation}
we insert a reservoir bit in prepared state $\ket{0}$, yielding the four state
density matrix
\begin{equation}
\rho=
\rightarrow
	\left( \begin{array}{cccc}
      \rho_{0 0} & \rho_{0 1}& 0 &
      0  \\
\rho_{1 0} & \rho_{1 1}  & 0 & 0 \\
0 &0 & 0 & 0 \\
0 &0 & 0 & 0 \\
  \end{array}
\right),
\end{equation}
then entangle the system and reservoir bits using $\CNOT^{s
\rightarrow r}$, yielding 
\begin{equation}
\rho=\left( \begin{array}{cccc}
      \rho_{0 0} & 0 & 0 &
      \rho_{0 1}  \\
0 &0 & 0 & 0 \\
0 &0 & 0 & 0 \\
\rho_{1 0} & 0 & 0 &
      \rho_{1 1}  \\
  \end{array}
\right).
\end{equation}
Next, the reservoir bit is used to control the evolution of the system bit
according to $\CR^{r \rightarrow s}(2 \theta)$, yielding 
\begin{equation} 
  \rho=
\left(
\begin{array}{cccc}
 \rho_{0 0} & 0 & \rho_{0 1} \sin (\theta ) & {\rho_{0 1}} \cos (\theta ) \\
 0 & 0 & 0 & 0 \\
 {\rho_{1 0}} \sin (\theta ) & 0 & {\rho_{1 1}} \sin ^2(\theta ) & {\rho_{1 1}} \cos (\theta
   ) \sin (\theta ) \\
 {\rho_{1 0}} \cos (\theta ) & 0 & {\rho_{1 1}} \cos (\theta ) \sin (\theta ) & {\rho_{1 1}}
   \cos ^2(\theta ) \\
\end{array}
\right).
\end{equation}
The reservoir bit is deleted by nonselective measurement along the x axis,
where $\ket{\pm x}=(\ket{1} \pm \ket{0})\sqrt{2}$, to yield
\begin{equation}
  \rho=
\left(
\begin{array}{cccc}
 \frac{1}{2} \left({\rho_{1 1}} \sin ^2(\theta )+{\rho_{0 0}}\right) & 0 & \frac{1}{2}
   ({\rho_{0 1}}+{\rho_{1 0}}) \sin (\theta ) & 0 \\
 0 & \frac{1}{2} {\rho_{1 1}} \cos ^2(\theta ) & 0 & 0 \\
 \frac{1}{2} ({\rho_{0 1}}+{\rho_{1 0}}) \sin (\theta ) & 0 & \frac{1}{2} \left({\rho_{1 1}}
   \sin ^2(\theta )+{\rho_{0 0}}\right) & 0 \\
 0 & 0 & 0 & \frac{1}{2} {\rho_{1 1}} \cos ^2(\theta ) \\
\end{array}
\right).
\end{equation}
Finally, tracing over the state of the reservoir bit yields the reduced
two state density matrix for the system alone
\begin{equation}
	\rho=\left(
	\begin{array}{cc}
		\rho_{0 0} + \rho_{1 1} \sin^{2}(\theta)   & 0 \\
		0 & \rho_{1 1} \cos^{2}(\theta) 
	\end{array}
	\right).
	\label{eq:two_state_after_pop_transfer}
\end{equation}

Note that by construction the transfer of population depends on the initial
population of $\alpha'$ and the rotation angle, but not on the off-diagonal
coherence terms $\rho_{\alpha \alpha'}$ or $\rho_{\alpha' \alpha}$.  Before the
trace, the two contributions to the $s=0$ population are $\ket{s,r}=\ket{0,0}$,
reflecting population which began in state 0, and $\ket{s,r}=\ket{0,1}$,
reflecting population which began in state 1 and was transferred to state 0 by
the controlled rotation operator.  Because these two states differ in the value
of the $r$ bit, they are summed incoherently by the trace over $r$, so that the
off diagonal terms do not contribute to the fraction of the population which is
transferred.


As constructed, the irreversible population transfer superoperator operates by
decimation.  A fixed fraction of the population of state $\ket{1}$ is
transferred to state $\ket{0}$ with every application of the superoperator,
with the fraction transferred determined by the choice of decimation angle
$\theta$.  For the specific application of depopulating state $\ket{1}$,
repeated application is not needed -- one may simply choose decimation angle
$\theta=\pi/2$ and completely depopulate the state.  However, as the
superoperator is extended to deal with logical problems, it will be more
convenient to treat $\theta$ as a small variable and depopulate unwanted states
by repeated decimation.

\begin{figure}[h!]
  \begin{center}
      \includegraphics[width=\columnwidth]{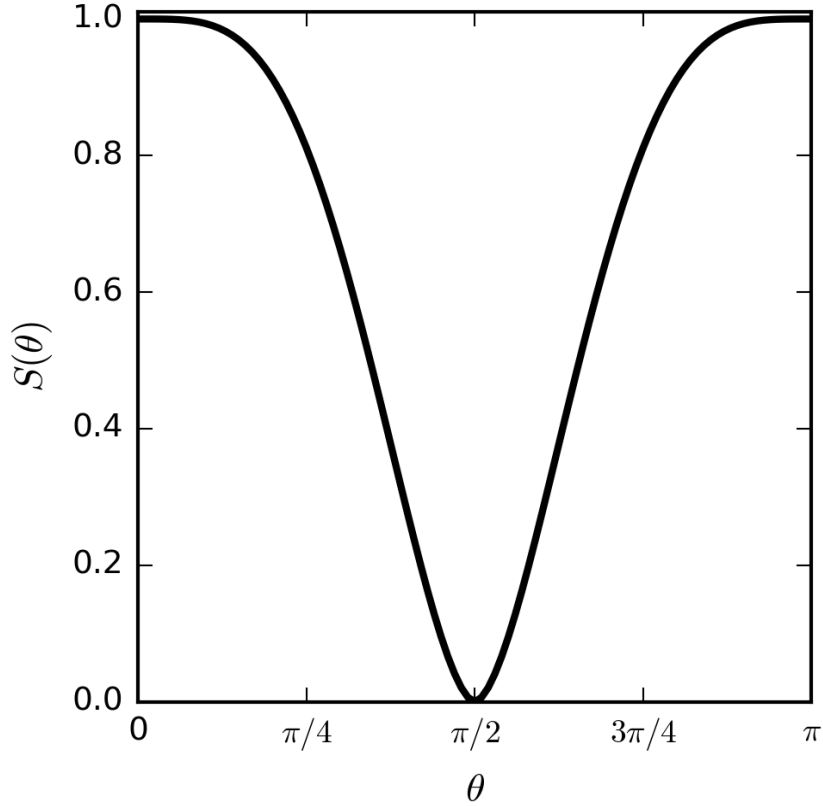}
  \end{center}
    \caption{Von Neumann entropy  $S$ vs decimation angle $\theta$ for
        irreversible population transfer from maximally disordered density
        matrix for two bits, as given by Eqs.
        \ref{eq:two_state_after_pop_transfer} and \ref{eq:vonneumanns}.
        $\theta=\pi/2$ transfers all population from state $\ket{1}$ to state
        $\ket{0}$, leaving the system in a zero entropy pure state.  
        }
  \label{fig:s_vs_theta}
\end{figure}

\paragraph{Change of entropy due to population transfer}
As seen in Eq. \ref{eq:two_state_after_pop_transfer}, the incoherent population transfer superoperator
transfers population fraction $\sin^{2}(\theta)$ along the diagonal of the
density matrix, regardless of the off diagonal coherence terms $\rho_{01}$ and
$\rho_{10}$.  In general, this superoperator is not invertible, and will change
the Von Neumann entropy of the system's reduced density matrix.  This
contrasts with operators in the Deutsch model which, being unitary, leave the
Von Neumann entropy of the system unchanged. Figure \ref{fig:s_vs_theta} shows the Von Neumann
entropy for a two state system beginning in state $\rho=\left(
\begin{array}{cc} 1/2 & 0 \\ 0 & 1/2 \end{array} \right)$ and acted upon by
the incoherent population transfer superoperator with angle $\theta$.
As the system begins in the maximally disordered state, the entropy
must decrease or stay the same.  $\theta=\pi/2$ transfers all population
from state $\ket{1}$ to state $\ket{0}$, leaving a pure state with
entropy 0.

\subsection{Logical Error Correction Superoperators}
\label{section:lec}

The irreversible population transfer superoperator defined in Eq.
\ref{eq:IPT} can be converted into an error correction
superoperator by making two changes.  First, $CNOT^{s \rightarrow
r}$ is replaced by an operator $C^{s \rightarrow r}$, which has the property that
\begin{equation}
	C^{s \rightarrow r}\ket{\alpha,0} = \ket{\alpha,1}
\end{equation}
for every state $\ket{\alpha}$ failing C and
\begin{equation}
	C^{s \rightarrow r}\ket{\alpha',0} = \ket{\alpha',0}
\end{equation}
for every state $\ket{\alpha'}$ satisfying C.  Here the combined state of the
system $s$ and reservoir $r$ are written $\ket{s,r}$.

Second, $\sigma_{y}^{t}$ in Eq.
\ref{eq:CR} is replaced by 
\begin{equation}
	\sigma_{y} \rightarrow \Sigma_{y} = \sum_{\alpha} i
	(\ket{\alpha}\bra{\alpha'(\alpha)} -
	\ket{\alpha'(\alpha)}\bra{\alpha}),
\end{equation}
so that every state $\ket{\alpha}$ failing clause $C$ is rotated into some
state  $\ket{\alpha'(\alpha)}$ that satisfies it.  As with the irreversible
population transfer superoperator, repeated application of $E^{(C)}(\theta)$
reduces population of failing states by decimation, transferring a fraction
$\sin^{2}(\theta)$ of the state population away from the failing states with
each application.

This method of error correction differs from that of Shor \cite{shor1995scheme}
in being fully quantum mechanical, and in considering logical as well as
physical errors.  It is fully quantum mechanical in the sense that the
existence of an error is never selectively measured, and the operations
performed by the computer on the system are identical whether an error exists
or not.  Thus, there is no need to diagnose a (classical) error syndrome and
take corrective action.  The errors being corrected are logical rather than
physical, in the sense that any population of a state which fails the clause is
treated as an error and reduced, regardless of whether that population arises
due to initial conditions, decoherence processes, imperfect operation of the
computer, or simply as a byproduct of other parts of the algorithm.  This
contrasts with the approach of Shor, in which an error consists of a difference
between the state of a quantum system and the desired state which would result
from perfect operation of the computer in the absence of decoherence and has no
independent logical meaning.  In this way, the error correction superoperator
may be considered a type of fixed point quantum search\cite{grover2005fixed} \cite{tulsi2006newalgorithm}, in which the system always moves closer to the desired
end state.

\subsection{Application to 1SAT}
\label{sect:3SAT}

A trivial example of using logical error correction as a solution algorithm
consists of solving a logical problem $L=\cup_{i}C_{i}$  of the form
\begin{equation}
        C = (\neg^{n_{a}} X_{a}),
	\label{eq:1sat_clause}
\end{equation}
where $\neg^{n}$ is a logical $\NOT$ raised to the power of n, $n_{i}$ may be 0
or 1, and each clause in the problem consists of a single variable.

The irreversible population transfer superoperator of Eq. \ref{eq:IPT} can be
made into an error correction superoperator for Eq. \ref{eq:1sat_clause} by
inserting $(\neg^{r})^{1-n_{i}}$ between the controlled not and the controlled
rotation to yield an error correction superoperator
\begin{equation}
        E^{s}(\theta) = \Tr_{r}\mathfrak{N}^{r}_{x}\CR^{r \rightarrow s}(2 \theta) (\neg^{r})^{(1-n_{i})} \CNOT^{s \rightarrow
	r} \mathfrak{I}_{ \ket{r}=\ket{0}},
	\label{eq:DEC}
\end{equation}
which reduces the population of any state having the wrong value in bit $i$ by
a factor of $\cos^{2}(\theta)$ on every application.

Choosing $\theta=\pi/2$, it can be seen that every application of the
decimation superoperator to a new clause concentrates probability into half as
many states as before, thereby reducing the system entropy by one bit.
Applying the decimation to all $N$ clauses concentrates probability from all
$2^{N}$ initial states into the same final state, which by virtue of never
failing a clause must be the solution.

Although it is certainly overkill to use a quantum computer to solve a problem
as trivial as 1SAT, this algorithm illustrates two generic features of
algorithms based upon logical error correction.  First, each application of a
decimation superoperator changes the probabilities for all $2^{N}$ states in
the search space.  Second, the effect of applying several decimation
superoperators in succession is to concentrate probability from exponentially
many initial states into a small number of states which are determined by the
logical criteria of satisfying each clause in the problem.  Both of these
features carry over to the significantly more interesting problem of 3SAT.

\subsection{Application to 3SAT}

A problem of particular interest in computer science is the solution of
Boolean formulae containing three variables per clause, or 3SAT.  
Here the logical problem $L=\cup_{i}C_{i}$ is a set of three variable clauses
of the form
\begin{equation}
	C = (\neg^{n_{a}} X_{a} \lor \neg^{n_{b}} X_{b} \lor \neg^{n_{c}}
	X_{c}),
	\label{eq:clause}
\end{equation}
where the clauses now contain three variables rather than one.  3SAT has the
property of being NP complete\cite{karp1972reducibility}, so that any problem
whose answer can be checked for correctness in polynomial time can be reduced
to 3SAT in polynomial time as well.  Because of this, an efficient algorithm
for solving 3SAT doubles as an efficient solution algorithm for all problems in
class NP, including many that cannot be solved efficiently using a classical
computer using known techniques.

It can be seen that 3SAT has the logical structure defined in section
\ref{section:lec}, where the three variable clauses are interpreted as error
criteria.  Noting that only one out of eight possible states for the three
variables in a given clause actually fails the clause, it follows that toggling
the states of any one-, two-, or three variables in the clause converts a state
that fails the clause into one that satisfies it.  From an entropic
perspective, concentrating population from eight initial states into the seven
final states which satisfy a given clause may be expected to decrease the
entropy by approximately $\log(7/8)/log(2) = -0.192$ bits.

An error correction superoperator for clause $C$ can be defined by
replacing $CNOT^{s \rightarrow r}$ in Eq. \ref{eq:IPT} with

    \begin{equation}
        C^{X_{a}, X_{b}, X_{c} \rightarrow R} = \neg_{a}^{n_{a}+1}
        \neg_{b}^{n_{b}+1} \neg_{c}^{n_{c}+1} T^{X_{a}, X_{b}, X_{c}
        \rightarrow r} \neg_{c}^{n_{c}+1} \neg_{b}^{n_{b}+1}
        \neg_{a}^{n_{a}+1},
        \label{eq:clauseC}
    \end{equation}
where $\neg_{a}$ flips the state of $X_{a}$ and $T^{X_{a}, X_{b}, X_{c}
\rightarrow r}$ is a generalized Toffoli gate which flips the state of the
target bit if all three inputs are true.  Here the specific choice of gates is
unimportant -- what is desired is that the reservoir bit encode the
satisfaction of the logical clause.  The controlled rotation $\CR^{R
\rightarrow S}(2 \theta)$ is replaced by three controlled rotations
\begin{equation}
	\CR^{R \rightarrow X_{a}}(2 \theta) \CR^{R \rightarrow X_{b}}(2 \theta)
	\CR^{R
		\rightarrow X_{c}}(2 \theta),
\end{equation}
so that to leading order, if state $\ket{s}$ fails the clause, a fraction of
the population equal to $\sin^{2}(\theta)$ is transferred from $\ket{s}$ to
each of the three states that differ from $\ket{s}$ by the value of one
variable in the clause.

Note that the transfer of population resulting from this superoperator differs
in an essential way from the classical algorithm of Sch\"oning
\cite{schoning1999probabilistic} and its quantum mechanical analogue
\cite{farhi2016quantum}.  In those algorithms, an error syndrome is selectively
measured, and corrective action is taken if the measurement returns an error.
(Since a classical system does not allow for superpositions, all classical
measurements are selective by default.)  By the rules of quantum measurement,
any measurement which returns an error projects all probability into the
subspace of error states, regardless of how much probability had accumulated in
solution states before the measurement was taken.  In contrast, the error
correction superoperator defined above does not make any selective
measurements, and allows probability to continue accumulating in the solution
states.

As with 1SAT, a solution algorithm consists of successively applying the error
correction superoperators corresponding to each clause in $L$.  However, here
the decimation angle $\theta$ will be treated as a free parameter rather than
being set to $\pi/2$.

\section{Running time analysis -- the decay of nonsolution probability}
\label{sect:running_time}

This section calculates the expected running time for the solution algorithm
defined in Section \ref{sect:3SAT}.  This is done by finding the rate of decay
for the nonsolution eigenvectors of the population transfer matrix which
results from applying the error correction superoperators in succession.  As
will be seen, the choice of decimation angles establishes a minimum rate of
decay for all nonsolution eigenvectors save one.  The nonsolution probability
contained in the remaining slowly decaying eigenvector may then be projected
into the quickly decaying subspace by varying the chosen decimation angles.

Varying the decimation angles in this way yields an exponential decay in the
probability of measuring a state which does not correspond to a valid solution.
As the number of gates required for a given error correction superoperator is
constant, the overall running time of the algorithm scales linearly with the
number of clauses in the problem.

\subsection{The population transfer matrix}

Whereas an error correction superoperator for a single clause may operate by
transferring population directly from states which fail that clause to states
which satisfy it, an error correction superoperator for an entire logical
problem must transfer population indirectly, by successively applying the error
correction superoperators corresponding to each clause in the problem.

Let a compound error correction superoperator for logical problem $L$ 
be constructed by combining the error correction superoperators for each
clause $C$ in $L$
\begin{equation}
        E^{(L)}(\vec{\theta})=E^{(C_{1})}(\theta_{1}) E^{(C_{2})}(\theta_{2}) E^{(C_{3})}(\theta_{3})
	\dots
\end{equation}
Any state that fails at least one clause will lose population due to the
corresponding $E^{(C_{i})}(\theta_{i})$, while solution states, which by
definition satisfy all clauses, act as population sinks.

Here the transfer of population from an initial nonsolution state to a final
solution state will in general be indirect, involving the transfer of
population between many intermediate nonsolution states.  The overall decay of
the nonsolution probability is given by the eigenvalue spectrum of the compound
error correction superoperator.

Because the error correction superoperators transfer population along the
diagonal of the density matrix, with no contribution from off-diagonal
coherence terms, the flow of population can be tracked by defining population
transfer matrices $T^{C_{i}}(\theta_{i})$ such that 
$T^{C_{i}}(\theta_{i})_{i \rightarrow j}$  
gives the population transferred from state $j$ to state $i$ by
$E^{(C_{i})}(\theta_{i})$.  To simplify the notation, the clause and decimation
angle for a given transfer matrix will be suppressed where this information can
be inferred from context.  The transfer matrix for the compound error
correction superoperator is now the matrix product of the transfer matrices for
the individual error correction superoperators.
\begin{equation}
	T(\vec{\theta}) = T^{C_{1}}(\theta_{1}) T^{C_{2}}(\theta_{2}) \dots,
\end{equation}
and the rate of decay for the nonsolution population is given by the
eigenvalues of the compound transfer matrix, which in turn depends
parametrically on the vector of decimation angles $\vec{\theta_{i}}$.

Expanding the initial population in eigenvectors of the transfer matrix
\begin{equation}
    \vec{P}(0) =\sum_{V}A_{V} \vec{V},
\end{equation}
where
\begin{equation}
    T(\vec{\theta}) \vec{V} = \lambda_{V} \vec{V},
\end{equation}
for each eigenvector $V$, and $A_{V}$ are expansion coefficients,  the state
vector after $N$ applications of the composite transfer matrix is given by
\begin{equation}
	P_{\alpha}(N) =\sum_{V} \lambda_{V}^{N} A_{V} \vec{V}.
	\label{eq:Pn}
\end{equation}

By construction, each solution state corresponds to an eigenvector of
$T(\vec{\theta})$ with eigenvalue 1, while the requirements that the population
of every state be finite and nonnegative and that all probabilities must sum to
1 mean that $\lambda_{V} \le 1$ for all $V$.

%

\subsection{Decay of nonsolution eigenvectors}

Provided that at least one solution state $\ket{\alpha*}$ exists, it is
straightforward to verify that all nonsolution eigenvectors must have nonzero
decay rates.  Consider nonsolution eigenvector $V$, which contains population
in at least one nonsolution state $\ket{\alpha}$.  Because $\ket{\alpha}$ is a
nonsolution state, it must fail at least one clause $C$.  Because
$\ket{\alpha*}$ satisfies $C$, it must differ from $\ket{\alpha}$ in at least
one of $X_{C_{1}}$, $X_{C_{2}}$, or $X_{C_{3}}$, the three variables in clause
C. The error correction superoperator $E^{(C)}(\theta)$  transfers population
to each of three states which differ from $\ket{\alpha}$, at least one of which
must have a Hamming distance to $\ket{\alpha*}$ smaller than the distance
between $\ket{\alpha*}$ and $\ket{\alpha}$.  Let this state be $\ket{\alpha'}$.
Now $\ket{\alpha'}$ must either be a solution state or fail some clause $C'$.
In the latter case, some population must be transferred to state
$\ket{\alpha''}$ which further decreases the Hamming distance to
$\ket{\alpha*}$.  For a problem with a finite number of variables, this chain
must eventually terminate with population being transferred to a solution
state, resulting in a nonzero decay rate for the nonsolution probability in
eigenvector $V$.

\subsection{The quasiequilibrium eigenvector}
\label{section:quasiequilibrium}

The eigenvalue spectrum for the composite transfer matrix $T(\vec{\theta})$
depends parametrically on the choice of decimation angles corresponding to the
individual clauses.  Here it can be shown that the smallest decimation angle
$\theta_{min}$ establishes a characteristic timescale for the decay of all
nonsolution eigenvectors save one.  Because $T(\vec{\theta})$ is real and
non-negative, this is a special case of the Perron Frobenius
theorem\cite{perron1907theorie} \cite{frobenius1912matrizen} \cite{pillai2005perron}.  For
the sake of notational clarity, the vector of decimation angles $\vec{\theta}$
will be suppressed for the remainder of this section.

The change in population for state s due to a single application of $T$ is given by
\begin{equation}
  \Delta P_{s}=\sum_{s'} T_{s' \rightarrow s} P_{s'} - T_{s \rightarrow s'}
  P_{s}.
  \label{eq:deltaP}
\end{equation}

Here it is convenient to treat the decimation angle $\theta_{i}$ for a
particular clause as a small parameter, so that the order in which the error
correction superoperators are applied may be neglected.  Noting that every
nonsolution state fails at least one clause and thereby loses a fraction
$\sin^{2}(\theta_{i})$ of its population, to leading order
\begin{equation}
	\sum_{s'} T_{s \rightarrow s'} \ge \theta_{\text{min}}^{2}
	\label{eq:lossfraction}
\end{equation}
for all nonsolution states s.  Thus, the minimum loss rate for a nonsolution
state is determined by the smallest decimation angle.

The condition that an eigenvalue decay slowly relative to this minimum loss
rate is sufficient to uniquely determine the state populations of the slowly
decaying eigenvector, up to normalization.
Writing Eq. \ref{eq:deltaP} as an eigenvalue equation
\begin{equation}
  \sum_{s'} T_{s' \rightarrow s} P_{s'} = \sum_{s'} T_{s \rightarrow s'} P_{s} - \nu
  P_{s},
	\label{eq:eigenvalue_linearsystem}
\end{equation}
if $\nu << \theta_{\text{min}}^{2}$ it can be treated as a small variable and
neglected, leaving a linear system with as many equations as variables  
\begin{equation}
  \sum_{s'} T_{s' \rightarrow s} P_{s'} = \sum_{s'}  T_{s \rightarrow s'}
	P_{s} .
  \label{eq:nosolution_equilibrium}
\end{equation}

Thus, regardless of the choice of transfer matrix, there will always be one
slowly decaying nonsolution eigenvector.  Because this eigenvector is is in
near equilibrium, it will also be referred to as the quasiequilibrium
eigenvector.  The remaining nonsolution eigenvectors, which do not satisfy Eq.
\ref{eq:nosolution_equilibrium}, decay quickly relative to this rate and will
be referred to as the quickly decaying subspace.

Applying this to Eq. \ref{eq:Pn}, all nonsolution probability contained in the
quickly decaying subspace will be rapidly transferred to the solution states at
a rate determined by $\theta_{min}$, which is chosen by the user.   In
contrast, the original population contained in the quasiequilibrium eigenvector
will decay slowly.  After repeated application of T, the remaining nonsolution
probability will preferentially be contained in the slowly decaying
eigenvector, and the decay of nonsolution probability will stagnate.



\subsection{Accelerating the decay of the quasiequilibrium probability
distribution}


The slow rate of decay for the quasiequilibrium eigenvector means that the
worst case behavior for the compound error correction superoperator is much
worse than the average case.  The quickly decaying subspace contains $O(2^{N})$
eigenvectors, compared to the quasistatic eigenvector's one.   If the
eigenvectors were populated at random, the bulk of the nonsolution probability
would be in the quickly decaying subspace, and would decay at a rate comparable
to $\theta^{2}_{min}$.  However, the nonsolution population contained in the
quasistatic eigenstate would decay at a rate much smaller than this and
approaching zero.


The worst case performance can be greatly accelerated by altering the choice
of decimation angles corresponding to particular clauses.  Because the
quasiequilibrium eigenvector depends parametrically on the choice of
decimation angles, altering these angles will project the slowly decaying
eigenvector of the old transfer matrix onto the quickly decaying eigenspace of
the new transfer matrix.

The worst case performance can be greatly accelerated by periodically altering
the decimation angles for particular clauses.  This changes the transfer matrix
$T(\vec{\theta})$, which by Eq. \ref{eq:nosolution_equilibrium} changes the
quasistatic eigenvector as well.  The old quasistatic eigenvector is then
projected into the eigenvector space of the new transfer matrix.  The
nonsolution population which is projected onto the quickly decaying subspace of
the new transfer matrix is then quickly transferred to the solution states.
Repeatedly varying the decimation angles in this way results in an exponential
decrease in the population contained in the quasistatic eigenvector.  This
section describes a simple ``back and forth'' iteration, in which the
decimation angles for different clauses are randomly varied, then changed back
to their original values.  This results in an exponential decay in the
nonsolution probability contained in the quasistatic eigenvector, at a rate
which is chosen by the user.

Let $\vec{V}^{(1)}$ and $\vec{W}_{n}^{(1)}$ be the slowly-
and quickly decaying eigendistributions corresponding to transfer matrix
$T^{(1)}(\vec{\theta}_{1})$ and $\vec{V}^{(2)}$ and $\vec{W}_{n}^{(2)}$ be the slowly-
and quickly decaying eigendistributions corresponding to transfer matrix
$T^{(2)}(\vec{\theta}_{2})$. 
Expand $V^{(1)}$ in the eigenbasis of $T^{(2)}$ to get
\begin{equation}
    \vec{V^{(1)}} = B_{V}  \vec{V}^{(2)} + \sum_{n} B_{n} \vec{W}_{n}^{(2)}
\end{equation}
and $\vec{V}^{(2)}$ in the eigenbasis of $T^{(1)}$ to get
\begin{equation}
    \vec{V^{(2)}} = C_{V}  \vec{V}^{(1)} + \sum_{n} C_{n} \vec{W}_{n}^{(1)}.
\end{equation}
Acting with $T^{(2)}$ on $\vec{V}^{(1)}$ yields
\begin{equation}
	T^{(2)} \vec{V}^{(1)}= (1+a_{V}) \vec{V}^{(2)} + \sum_{n} A_{n} \vec{W}^{(2)}_{n},
	\label{eq:blowoff_expansion}
\end{equation}
where $a_{V}=A_{V}-1$.

Beginning with all nonsolution probability in the slowly decaying
eigenstate $\vec{V}^{(1)}$ and
changing from $T^{(1)}$ to $T^{(2)}$ projects a fraction $(1-B_{V})$ of
the nonsolution probability into the quickly decaying subspace, where
repeated iteration will rapidly transfer it to the solution states,
leaving the quasiequilibrium probability equal to $B_{V}\vec{V}^{(2)}$.  Changing
back to transfer matrix $T^{(1)}$ now projects fraction $(1-C_{V})$ of
the remaining population into the quickly decaying subspace, so that the
population of the slowly decaying subspace has decreased by a factor of
$B_{V}C_{V}$.

The coefficient $B_{V}C_{V}$ can be found to leading order by approximating the decay rate of the
slowly decaying eigendistributions as zero, so that
\begin{eqnarray}
    T^{(1)} \vec{V}^{(1)} =& \vec{V}^{(1)} \\
    T^{(2)} \vec{V}^{(2)} =& \vec{V}^{(2)}.
\end{eqnarray}
Let $\Delta T = T^{(2)} - T^{(1)}$ and $\Delta \vec{V} = \vec{V}^{(2)} - \vec{V}^{(1)}$, so that
Eq. \ref{eq:blowoff_expansion} can be rewritten
\begin{equation}
    a_{V} \vec{V}^{(2)} + \sum_{n} A_{n} \vec{W}^{(2)}_{n} = -T^{(2)} \Delta \vec{V}.
	\label{eq:blowoff_difference}
\end{equation}
Expanding the population difference 
\begin{equation}
    \Delta \vec{V} = B_{V} \vec{V}^{(2)} + \sum_{n} B_{n} \vec{W}^{(2)}_{n},
\end{equation}
in the eigenbasis of $T^{(2)}$, it follows that 
\begin{equation}
	a_{V} = - B_{V} = \frac{\Delta \vec{V} \cdot
	\vec{V}^{(2)}}{\vec{V}^{(2)}\cdot \vec{V}^{(2)}},
	\label{eq:blowoff_loss}
\end{equation}
so that the loss of probability is largest when $\vec{V}^{(1)}$ and
$\vec{V}^{(2)}$ are most dissimilar.

A simple scheme for transferring population out of the slowly decaying
eigenstate is to randomly vary the decimation angles corresponding to
particular clauses by a fixed amount.
Let $T^{(1)}=T(\theta_{0}, \theta_{0}, \dots)$ be the transfer matrix
corresponding to setting all decimation angles equal, and 
$T^{(2)}= T(\theta_{0}(1 + \eta \sigma_{1}), \theta_{0}(1 + \eta \sigma_{2}),
\dots)$, where $\sigma_{i}$ is randomly chosen to be plus or minus 1, so that
the decimation angle for each clause is randomly increased or decreased by a
fixed fraction.  Treating $\theta_{0}$ and $\eta$ as small parameters, to
leading order the fraction of the population transferred from a state failing
clause $C$ by $T^{(C)}$ is
\begin{equation}
	T^{(C)} = \theta_{C}^{2} = \theta_{0}^{2}(1+ 2
	\eta \sigma_{C}).
	\label{eq:newT}
\end{equation}

%
%

The difference between the slowly decaying eigenvectors for $T^{(1)}$ and
$T^{(2)}$ is now
\begin{equation}
    \Delta \vec{V}[s] = \sum_{C} \Delta \vec{V}^{(C)}[s] + \Delta \vec{V}^{(R)}[s],
	\label{eq:deltaV}
\end{equation}
where $\Delta \vec{V}^{(C)}$ are first order corrections offsetting changes in
$T^{(C)}$, while $\Delta \vec{V}^{(R)}$ are residual corrections enforcing that the
new quasiequilibrium distribution obeys Eq. \ref{eq:nosolution_equilibrium}.

To leading order, the change in population for state $s$ due to a small change
in the decimation angle is found by setting
\begin{equation}
	\sum_{s'} (\vec{V}[s] + \Delta \vec{V}^{(C)}[s])(T^{(C)}_{s' \rightarrow s}+
	\Delta T^{(C)}_{s' \rightarrow s}) = \sum_{s'} \vec{V}[s] T^{(C)}_{s'
	\rightarrow s},
\end{equation}
so that after neglecting the doubly small term $\Delta T^{(C)}_{s' \rightarrow
s} \Delta \vec{V}^{(C)}[s]$,
\begin{equation}
	\frac{\Delta \vec{V}^{(C)}[s]}{\vec{V}[s]} = - \frac{\sum_{s'} \Delta
		T^{(C)}_{s' \rightarrow s}} {T^{(C)}_{s' \rightarrow s}} = - 2 \eta
		\sigma_{C}.
\end{equation}
Subtracting these terms from Eq. \ref{eq:deltaV} yields a linear system for the
remaining corrections

\begin{equation}
        \begin{split}
	\sum_{s', C} (\Delta \vec{V}^{R}[s] T^{(C)}_{s \rightarrow s'} -
                \Delta \vec{V}^{R}[s'] T^{C}_{s' \rightarrow s})= \\
                \sum_{s', C' \ne C} (\Delta \vec{V}^{(C')}[s'] T^{(C)}_{s' \rightarrow s} -
	\Delta \vec{V}^{(C')}[s] T^{(C)}_{s \rightarrow s'}),
	\label{eq:deltaPr_linearsystem}
        \end{split}
\end{equation}
where the right hand side has expectation value 0.

The second step of the back and forth iteration changes all decimation angles
back to their original values, so that 
\begin{eqnarray}
\Delta T^{(C)} \rightarrow - \Delta T^{(C)}
\end{eqnarray}
and
\begin{equation}
	\Delta \vec{V} \rightarrow -\Delta \vec{V},
\end{equation}
and the slowly decaying eigendistribution $\vec{V}^{(2)}$ is projected into the
eigenbasis of $T^{(1)}$.

Both changes to the transfer matrix result in a loss of total nonsolution
probability, as population is transferred from the slowly decaying
to the quickly decaying eigendistributions.  The coefficient of the slowly decaying
eigendistribution after the second transfer is given by 
\begin{equation}
	\frac{A'_{V}}{A_{V}} = (1- \frac{\Delta \vec{V} \cdot
	\vec{V}^{(2)}}{\vec{V}^{(2)}\cdot \vec{V}^{(2)}})(1+ \frac{\Delta \vec{V} \cdot
	\vec{V}^{(2)}}{\vec{V}^{(2)}\cdot \vec{V}^{(2)}}).
\end{equation}

Let 
\begin{equation}
	\alpha = \frac{\sum_{C, s} \Delta \vec{V}[s] \vec{V}[s] }{ \sum_{s} \vec{V}[s] \vec{V}[s]}
	= \frac{\sum_{C} \sigma_{C} \eta \sum_{s} \vec{V}[s] \vec{V}[s]} { \sum_{s} \vec{V}[s] \vec{V}[s]}
\end{equation}
and
\begin{equation}
	\beta= \frac{\sum_{s} \Delta \vec{V}^{R}[s] \vec{V}[s]} { \sum_{s}
	\vec{V}[s] \vec{V}[s]},
\end{equation}
so that 
\begin{equation}
	\frac{A'_{V}}{A_{V}} = 1-\alpha^{2}-\alpha\beta-\beta^{2}.
	\label{eq:ratio_alphabeta}
\end{equation}
Then
\begin{equation}
	\alpha^{2} = \frac{\sum_{C, C'} \sigma_{C} \sigma_{C'} \eta^{2} \sum_{s, s'}
	\vec{V}[s] \vec{V}[s']} {\sum_{s, s'}
	\vec{V}[s] \vec{V}[s']}.
\end{equation}
Because $\sigma_{C}$ and $\sigma_{C'}$ are independent random variables with mean
value 0, 
\begin{equation}
	\sum_{C} \sigma_{C} =0
\end{equation}
and
\begin{equation}
	\sum_{C, C'} \sigma_{C}^{s} \sigma_{C'}^{s} =\sum_{C} \sigma_{C}^{2} =N_{C}^{s}
	\label{eq:clausesum}
\end{equation}
where $N_{C}^{s}$ is the number of clauses failed by state s, so that
$N_{C}^{s} \ge 1$ for every nonsolution state.
From this it follows that the expectation value
\begin{equation}
	\left<\frac{A'_{V}}{A_{V}}\right> \le (1 - \eta^{2}),
\end{equation}
so that the loss of population from the slowly decaying eigendistributions due to a
back and forth iteration cycle is controlled by the user controlled
parameter $\eta$.

\subsection{Expected Running Time}

The ``back and forth'' iteration scheme described in the previous section
yields a linear time solution algorithm for the 3SAT problem of interest.  If
\begin{equation}
	N = \log(\epsilon)/ \log(1-\theta_{0}^{2})
\end{equation}
and
\begin{equation}
	M = \log(\epsilon)/log(1-\eta^{2}),
\end{equation}
Then for an arbitrary initial probability vector $V_{0}$ containing initial nonsolution probability $P_{0}$, final population vector
\begin{equation}
	V_{f} = ((T^{(2)})^{N}(T^{(1)})^{N})^{M} V_{0}
\end{equation}
will contain nonsolution probability $P_{f}\le \epsilon P_{0}$.

Writing the number of operations required for a single error correction
superoperator as $N_{E}$ and the number of clauses in logical problem $L$ as
$N_{C}$, the expected running time for the algorithm is
\begin{equation}
    \text{Number of operations} =N_{E} N_{C} M N,
\end{equation}
which scales linearly with the number of clauses in the problem.

\section{Conclusions}

The algorithm presented in this paper is the first known algorithm for solving
an NP complete problem in sub-exponential time.  Provided they can be built,
irreversible quantum processors containing even a modest number of logical
qubits will be capable of solving a large class of computational problems
hitherto considered intractable for even the largest computational resources --
quantum or classical.  Moreover, by constructing error correction
superoperators which do not rely on contributions from off-diagonal coherence
terms, this algorithm reduces a significant technological barrier to the
construction of practical quantum processors.

An interesting question which is raised by this work is the extent to which
irreversible operations are necessary for solving computational problems
involving a large search space.  In Eq. \ref{eq:volfrac}, is there some
fundamental physical limit on the degree to which $f_{v}$ can be increased by
use of reversible operations alone?  If this were true, there might exist some
computational problems which require some minimum entropy to be removed from
the system before it can be solved in polynomial time using a reversible
quantum computer.  The use of thermodynamic arguments to constrain the behavior
of an arbitrary reversible process might serve as a new avenue to the still
unresolved question of P vs NP.

\appendix
\section{A simple example of back and forth iteration}

A simple example of back and forth iteration is obtained by solving a three
variable 3SAT problem in which $\ket{TTT}$ is the only solution.  This
corresponds to performing error correction operations for clauses 0-6 of Table
\ref{table:clauseorder}.  Each state other than $\ket{TTT}$ fails a single
clause.  The error correction superoperator corresponding to that clause then
transfers population away from the failing state to each of the three states
which differ from the failing state in a single variable.  Choosing
$\theta_{C}=\theta_{0}=0.1$ for all clauses and including terms to leading
order in $\theta$ yields transfer matrix
\begin{equation}
	T^{(1)} = \begin{array}{llllllll}
	0.97	&	0.01	&	0.01	&	0.0	&	0.01	&	0.0	&	0.0	&	0.0 \\
	0.01	&	0.97	&	0.0	&	0.01	&	0.0	&	0.01	&	0.0	&	0.0 \\
	0.01	&	0.0	&	0.97	&	0.01	&	0.0	&	0.0	&	0.01	&	0.0 \\
	0.0	&	0.01	&	0.01	&	0.97	&	0.0	&	0.0	&	0.0	&	0.0 \\
	0.01	&	0.0	&	0.0	&	0.0	&	0.97	&	0.01	&	0.01	&	0.0 \\
	0.0	&	0.01	&	0.0	&	0.0	&	0.01	&	0.97	&	0.0	&	0.0 \\
	0.0	&	0.0	&	0.01	&	0.0	&	0.01	&	0.0	&	0.97	&	0.0 \\
	0.0	&	0.0	&	0.0	&	0.01	&	0.0	&	0.01	&	0.01	&	1.0 \\
	\end{array}
\end{equation}
where the state ordering is taken from table \ref{table:clauseorder}.
The eigenvalues and eigendistributions of this transfer matrix are given by Table
\ref{table:T1eigs}

\begin{table*}
	\begin{tabular}{llll}
		\#	& Clause & State Decimated  & States Populated \\
		0 &	$(X_{1} \lor X_{2} \lor X_{3})$ & $\ket{FFF}$ &	$\ket{FFT}$, $\ket{FTF}$,
		$\ket{TFF}$	\\

		1 &	$(X_{1} \lor X_{2} \lor \neg X_{3})$ & $\ket{FFT}$ &	$\ket{FFF}$,
		$\ket{FTT}$, $\ket{TFT}$	\\

		2 &	$(X_{1} \lor \neg X_{2} \lor X_{3})$ & $\ket{FTF}$ &	$\ket{FTT}$,
		$\ket{FFF}$, $\ket{TTF}$	\\

		3 &	$(\neg X_{1} \lor \lnot X_{2} \lor \lnot X_{3})$ & $\ket{FTT}$ &
		$\ket{FTF}$, $\ket{FFT}$, $\ket{TTT}$	\\

		4 &	$(\lnot X_{1} \lor  X_{2} \lor  X_{3})$ & $\ket{TFF}$ &	$\ket{TFT}$,
		$\ket{TTF}$, $\ket{FFF}$	\\

		5 &	$(\lnot X_{1} \lor  X_{2} \lor \lnot X_{3})$ & $\ket{TFT}$ &
		$\ket{TFF}$, $\ket{TTT}$, $\ket{FFT}$	\\

		6 &	$(\lnot X_{1} \lor \lnot X_{2} \lor X_{3})$ & $\ket{TTF}$ &	$\ket{TTT}$,
		$\ket{TFF}$, $\ket{FTF}$	\\

		7 &	$(\neg X_{1} \lor \neg X_{2} \lor \neg X_{3})$ & $\ket{TTT}$ &
		$\ket{TTF}$, $\ket{TFT}$, $\ket{FTT}$	\\

	\end{tabular}
	\caption{Each of the eight three variable states fails a single clause.  To
		leading order, the error correction operator for this clause depopulates the state
		which fails the clause and populates the three states which differ from
		this state by a single variable.
	}
	\label{table:clauseorder}
\end{table*}

\begin{table*}[h]
	\begin{tabular}{lll}
		Eigendistribution & $\lambda$ & $\vec{\lambda}$ \\
		$\vec{s^{*}}$	&	1.0	&	$ (0.0, 0.0, 0.0, 0.0, 0.0, 0.0, 0.0, 1.0)$ \\
		$\vec{V}^{(1)}$	&	0.996	&	$ (-0.165, -0.146, -0.146, -0.11, -0.146, -0.11, -0.11, 0.934)$ \\
		$\vec{W}^{(1)}_{1}$	&	0.98	&	$ (-0.0, 0.577, -0.289, 0.289, -0.289, 0.289, -0.577, 0.0)$ \\
		$\vec{W}^{(1)}_{2}$	&	0.98	&	$ (0.0, -0.086, -0.452, -0.537, 0.537, 0.452, 0.086, 0.0)$ \\
		$\vec{W}^{(1)}_{3}$	&	0.97	&	$ (0.707, 0.0, -0.0, -0.354, -0.0, -0.354, -0.354, 0.354)$ \\
		$\vec{W}^{(1)}_{4}$	&	0.96	&	$ (-0.0, -0.577, 0.289, 0.289, 0.289, 0.289, -0.577, 0.0)$ \\
		$\vec{W}^{(1)}_{5}$	&	0.96	&	$ (0.0, 0.128, -0.551, 0.424, 0.424, -0.551, 0.128, 0.0)$ \\
		$\vec{W}^{(1)}_{6}$	&	0.944	&	$ (-0.457, 0.403, 0.403, -0.305, 0.403, -0.305, -0.305, 0.162)$ \\
\end{tabular}
\caption{Eigenvalues and eigendistributions of $T^{(1)}$. }
\label{table:T1eigs}
\end{table*}

Setting $\eta=0.2$ and $\vec{\sigma}=(1,1,1,1,-1,-1,-1)$ changes the
transfer matrix to
\begin{equation}
	T^{(2)} =
	\begin{array}{llllllll}
	0.957	&	0.014	&	0.014	&	0.0	&	0.006	&	0.0	&	0.0	&	0.0	 \\
	0.014	&	0.957	&	0.0	&	0.014	&	0.0	&	0.006	&	0.0	&	0.0	 \\
	0.014	&	0.0	&	0.957	&	0.014	&	0.0	&	0.0	&	0.006	&	0.0	 \\
	0.0	&	0.014	&	0.014	&	0.957	&	0.0	&	0.0	&	0.0	&	0.0	 \\
	0.014	&	0.0	&	0.0	&	0.0	&	0.981	&	0.006	&	0.006	&	0.0	 \\
	0.0	&	0.014	&	0.0	&	0.0	&	0.006	&	0.981	&	0.0	&	0.0	 \\
	0.0	&	0.0	&	0.014	&	0.0	&	0.006	&	0.0	&	0.981	&	0.0	 \\
	0.0	&	0.0	&	0.0	&	0.014	&	0.0	&	0.006	&	0.006	&	1.0	 \\
	\end{array},
\end{equation}
with eigenvalues and eigendistributions given by Table \ref{table:T2eigs}.

\begin{table*}[h]
	\begin{tabular}{lll}
		Eigenstate	&	$\lambda$	& $\vec{\lambda}$ \\
		$\vec{s^{*}}$	&	1.0	&	$ (0.0, 0.0, 0.0, 0.0, 0.0, 0.0, 0.0, 1.0)$ \\
		$\vec{V}^{(2)}$	&	0.997	&	$ (-0.101, -0.088, -0.088, -0.064, -0.234, -0.175, -0.175, 0.924)$ \\
		$\vec{W}^{(2)}_{1}$	&	0.984	&	$ (0.0, -0.161, 0.161, -0.0, -0.0, -0.689, 0.689, 0.0)$ \\
		$\vec{W}^{(2)}_{2}$	&	0.98	&	$ (-0.205, -0.322, -0.322, -0.399, 0.702, 0.19, 0.19, 0.166)$ \\
		$\vec{W}^{(2)}_{3}$	&	0.974	&	$ (0.233, 0.02, 0.02, 0.034, 0.534, -0.546, -0.546, 0.25)$ \\
		$\vec{W}^{(2)}_{4}$	&	0.955	&	$ (0.617, 0.043, 0.043, -0.654, -0.377, 0.069, 0.069, 0.189)$ \\
		$\vec{W}^{(2)}_{5}$	&	0.953	&	$ (0.0, -0.626, 0.626, -0.0, -0.0, 0.329, -0.329, 0.0)$ \\
		$\vec{W}^{(2)}_{6}$	&	0.926	&	$ (-0.492, 0.482, 0.482, -0.457, 0.165, -0.147, -0.147, 0.115)$ \\
	\end{tabular}
	\caption{Eigenvalues and eigendistributions of $T^{(2)}$.}
\label{table:T2eigs}
\end{table*}

Note that in both Table \ref{table:T1eigs} and Table \ref{table:T2eigs}, the
decay rate for $\vec{V}^{(1)}$ and $V^{(2)}$ is small relative to
$\theta_{0}^{2}=0.01$, while the decay rates for eigendistributions $\vec{W}_{n}$
are comparable to this value or larger.  

Changing the transfer matrix from $T^{(1)}$ to $T^{(2)}$ projects $V^{(1)}$
into the new eigenbasis
\begin{equation}
    \begin{split}
        \vec{V^{(1)}} =& 1.005 \vec{V}^{(2)} + 0.163 \vec{W}^{(2)}_{2} - 0.085 \\
        &\vec{W}^{(2)}_{3} -0.028 \vec{W}^{(2)}_{4} - .005 \vec{W}^{(2)}_{6}.
    \end{split}
\end{equation}
Once the coefficients of the quickly decaying eigendistributions $\vec{W}_{n}$ have
decayed to zero, returning to the original transfer matrix projects $V^{(2)}$
into the original eigenbasis
\begin{equation}
    \begin{split}
        \vec{V}^{(2)} =& 0.956 \vec{V}^{(1)} + 0.055 \vec{W}^{(1)}_{1} - 0.13
\vec{W}^{(1)}_{2} + 0.085 \vec{W}^{(1)}_{3} \\
        &- 0.014 \vec{W}^{(1)}_{4} - 0.018
\vec{W}^{(1)}_{5} 0.007 \vec{W}^{(1)}_{6},
    \end{split}
\end{equation}
so that the coefficient of $\vec{V}^{(1)}$ after the back and forth iteration
is $1.005*.956=0.96\approx1-\eta^{2}$.

\section*{Funding}
This work was performed under the auspices of the U.S. Department of Energy by
Lawrence Livermore National Laboratory under Contract DE-AC52-07NA27344,
release number LLNL-JRNL-748132.

This document was prepared as an account of work sponsored by an agency of the
United States government. Neither the United States government nor Lawrence
Livermore National Security, LLC, nor any of their employees makes any
warranty, expressed or implied, or assumes any legal liability or
responsibility for the accuracy, completeness, or usefulness of any
information, apparatus, product, or process disclosed, or represents that its
use would not infringe privately owned rights. Reference herein to any specific
commercial product, process, or service by trade name, trademark, manufacturer,
or otherwise does not necessarily constitute or imply its endorsement,
recommendation, or favoring by the United States government or Lawrence
Livermore National Security, LLC. The views and opinions of authors expressed
herein do not necessarily state or reflect those of the United States
government or Lawrence Livermore National Security, LLC, and shall not be used
for advertising or product endorsement purposes.

\section*{Statement of Financial Interest} 
The contents of this paper have been incorporated into pending US patent
application number 2017/007-6220 and pending international patent application
number PCT/US2016/051376.  These applications predate the author's employment
with Lawrence Livermore National Laboratory.

\bibliographystyle{plain}

\end{document}